\documentclass[journal=jacs,manuscript=article]{achemso}
\usepackage{chemformula} % Formula subscripts using \ch{}
\usepackage[T1]{fontenc} % Use modern font encodings
\usepackage{graphicx}
\usepackage{amsmath}
\usepackage{amssymb}
\usepackage{hyperref}
\usepackage{color}
\usepackage[version=4]{mhchem}

\author{Xinyu Yang}
\author{Shuai Dong}
\email{sdong@seu.edu.cn}
\affiliation[Southeast University]
{Key Laboratory of Quantum Materials and Devices of Ministry of Education, School of Physics, Southeast University, Nanjing 211189, China}

\title[An \textsf{achemso} demo]
{Ferroelectric switchable altermagnetic-like compensated ferrimagnets with charge ordering}

\begin{document}

\begin{abstract}
Unconventional collinear magnets with almost zero magnetization but prominent nonrelativistic spin-splitting, such as altermagnets, can inherit the advantages of both ferromagnets and antiferromagnets. By incorporating more degrees of freedom such as ferroelectricity and charge ordering, these unconventional magnets can be even more interesting and functionalized. With this design principle, the Fe$_3$O$_5$ monolayer is predicted to exhibit a hybrid spin-splitting mechanism, with the superposition of the altermagnetic-like $k$-path alternating splitting and ferrimagnet-like Zeeman splitting. Benefiting from the hidden magnetoelectricity based on the spin-charge coupling, such spin-splitting can be fully switched by an electric field. Its conductivity is highly spin-polarized, with a polarization ratio above $99\%$, comparable to half-metals but with zero magnetization.
\end{abstract}


\section{Introduction}
\begin{figure}
	\includegraphics[width=0.5 \textwidth]{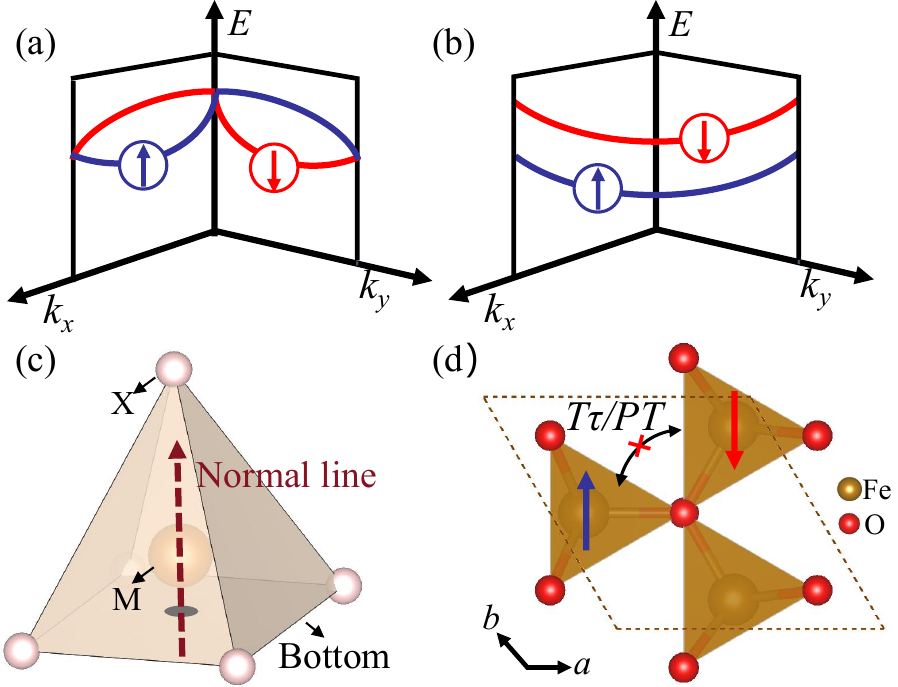}
	\caption{Schematic illustration of band splittings, including: (a) an altermagnetic-type $k$-dependent alternating splitting and (b) a ferrimagnet-type Zeeman splitting. (c) The unit structure of $X$-pyramid with a caged $M$ ion. (d) Top view of a primitive cell of $M_3X_5$ monolayer. Here $M$=Fe and $X$=O. $T\tau$ and $PT$ represent the operations of time-translation symmetry and parity-time reversal symmetry between sublattices with antiparallel spins respectively, both of which are broken here.}
	\label{F1}
\end{figure}

Unconventional collinear magnets, such as altermagnets and compensated ferrimagnets, exhibit nearly zero net magnetization but nonrelativistic spin splitting, thereby inheriting the advantages of ferromagnetism and antiferromagnetism. In altermagnets, the broken parity-time ($PT$) reversal symmetry and the conservation of mirror ($TM$) or rotational ($TR$) symmetry of antiparallel magnetic sublattices, give rise to alternating spin splitting in the momentum $k$-space, as illustrated in Figure~\ref{F1}(a) \cite{Libor2022prx031042,Libor2022prx040501, fender2025JACS,Ma2021nc2846,Hiroaki2019JPSJ123702,Liu2025PRL056801,yang2025AFMe17921,Liuprl2024206702,sun2025AM2502575,zhang2025JACS,Duan2025JACS,Che2026JACS,zhou2025prl106801,liu2025prl106802}. Determined by both the N\'eel order and the sublattice symmetry, spin-degeneration at certain high-symmetry $k$-points or along certain $k$-paths are still preserved by symmetry protection. In contrast, by breaking all symmetry operations between the spin-up and spin-down sites, the so-called compensated ferrimagnets exhibit Zeeman-type spin splitting in the entire Brillouin zone, as shown in Figure~\ref{F1}(b) \cite{yang2023apl162403,James2024prl216701,liu2025prl116703,zhang2025PRB024425,kan2025nl14960,zhao2025prb}. Recent theoretical studies have predicted that this Zeeman-type spin splitting can be realized in heterojunctions \cite{yang2023apl162403}, ordered Mn$_2$SiSnN$_4$ alloy \cite{James2024prl216701}, Janus bilayers \cite{liu2025prl116703}, and strained altermagnets \cite{kan2025nl14960,zhao2025prb}. However, all these approaches rely on fixed structural frameworks, and thus their Zeeman-type spin splittings are difficult to electrically switch.  

Ferroelectric control of magnetism is especially promising for designing energy-saving, high-density, and nonvolatile devices \cite{wang2003science1719,Kimura2003nature55,kan2025prl196801,Meng2025prl226402,wang2023pnase2305197120,wang2024nl,liu2026nl}. Recently, the compensated ferrimagnet GaFeO$_3$ was predicted to be electrically switchable by flipping its ferroelectric polarization \cite{zhang2025PRB024425}. However, its wide bandgap ($\sim2.72$ eV) and low carrier concentration make it not a preferred candidate for devices based on spin-dependent transport, and its three-dimensional perovskite structure lacks the advantages of either Si-based integration or electrostatic control at the nanoscale. Another solution is based on two-dimensional (2D) van der Waals (vdW) bilayers through interlayer-slide engineering, which can also create compensated ferrimagnetism \cite{wang2025prb195427,Liu2025PRL056801}. Nevertheless, there is another intrinsic limitation: the typically weak interlayer vdW couplings result in faint spin splittings on the order of merely several tens of meV, which cannot lead to a high ratio of spin polarization for electronic transport. Thus, it remains challenging to achieve nonvolatile electric-field control of prominent spin splittings in 2D compensated ferrimagnets.

Recently, $M_3X_5$ monolayers were predicted as a new branch of 2D materials, where $M$ denotes a transition metal and $X$ represents O, S, or Se \cite{xu2023apl163103,xu2023prb214427}. Their common structure can be derived from a sublayer of Kagome materials $AM_3X_5$ ($A$: alkaline earth) \cite{Toberer2019prm094407,gao2024nc9626}. Their structural characteristics are particularly unusual: each $M$ ion is enclosed within a square pyramidal $X$-cage (i.e. half of an octahedron), as shown in Figure~\ref{F1}(c). The $X$-pyramid is orientable, characterized by the normal line of the bottom. Interestingly, the orientations of neighboring pyramids are noncollinear, giving a $120^{\circ}$ configuration, as shown in Figure~\ref{F1}(d).
With such a structure, the time-translational ($T\tau$)  symmetry and $PT$ reversal symmetry are naturally broken between neighboring $M$-sites with an antiferromagnetic configuration, which may lead to nontrivial spin splittings. Moreover, by introducing additional degrees of freedom such as ferroelectricity and charge ordering, such spin splittings can be switched by tuning these symmetries.

Following this design principle, here we demonstrate that the Fe$_3$O$_5$ monolayer is multiferroic, exhibiting both ferroelectricity and altermagnetic-like compensated ferrimagnetism. Without spin-orbit coupling (SOC), its electronic band structure exhibits characteristics of both $k$-dependent alternating splitting and plain Zeeman splitting. As a narrow-gap ($\sim0.4$ eV) semiconductor, its spin-polarized conductivity is highly anisotropic as in altermagnets, with the maximum polarization ratio close to $100\%$, higher than most altermagnets. Such spin-polarized conductivity can be electrically switched, mediated via the spin-charge coupling magnetoelectricity \cite{Dong2019NSR}.

\section{Results and Discussion}
\begin{figure*}
	\includegraphics[width=0.98\textwidth]{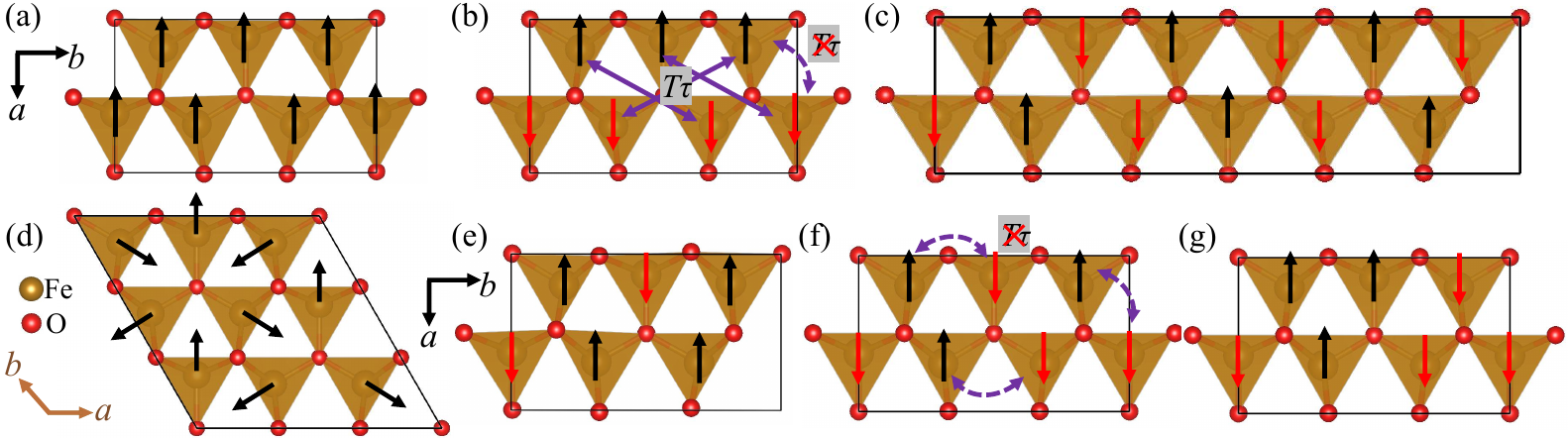}
	\caption{Schematic of seven most possible magnetic orders in Fe$_3$O$_5$ monolayer. (a) FM; (b) SAFM. (c) ZAFM; (d) YAFM; (e) UUD-ferrimagnetic (FiM); (f) UUD-AFM; (g) UUD-AFM2. Spins are denoted by on-site arrows. Dashed double-headed arrows: nearest-neighbor sites with opposite spins. Solid double-headed arrows: next-nearest-neighbor sites with opposite spins.}
	\label{F2}
\end{figure*}

For Fe's triangular lattice, there are four usual magnetic structures: ferromagnetic (FM), stripy-antiferromagnetic (SAFM), zigzag-antiferromagnetic (ZAFM), and $120^\circ$ noncollinear antiferromagnetic (YAFM), as shown in Figures~\ref{F2}(a-d). In addition, an exotic up-up-down (UUD) magnetic order was also reported in Ti$_3$O$_5$ and TmMgGaO$_4$ \cite{Philipp2020prx011007,xu2023prb214427}. Here three UUD variants: UUD-FiM, UUD-AFM, and UUD-AFM2, are considered, as shown in Figures~\ref{F2}(e-f).  

\begin{figure}
\includegraphics[width=0.6\textwidth]{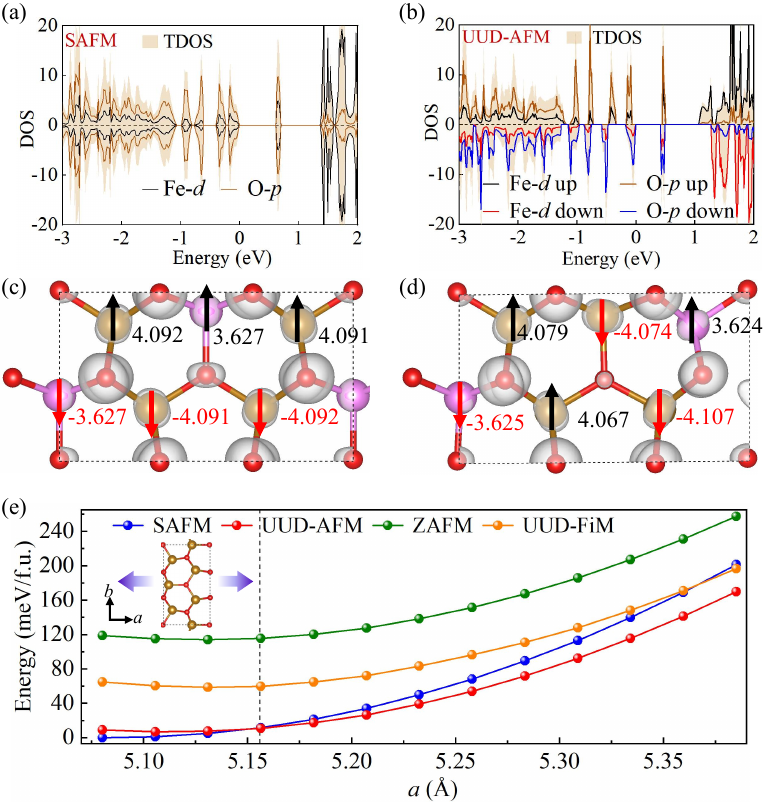}
\caption{Electronic structures of SAFM and UUD-AFM phases of Fe$_3$O$_5$ monolayer. (a-b) The total density of states (TDOS) and atomic-projected density of states (PDOS) of SAFM and UUD-AFM states. (c-d) Top views of valence electron densities (energy from $-1.5$ to $0$ eV) and local magnetic moments of Fe sites (within the default Wigner-Seitz radius). Fe$^{3+}$ and Fe$^{4+}$ sites are distinguished by colors. 
(e) Total energy of different magnetic orders as a function of lattice constant $a$. The lattice constant $b$ and atomic coordinates are relaxed correspondingly. The energy of the SAFM order is taken as the reference.}
\label{F3}
\end{figure}

The energies of all candidate states are calculated using density functional theory (DFT), as summarized in Table~\ref{Tab1}. Details of calculation methods can be found in Supporting Information (SI). The SAFM state has the lowest energy. The calculated lattice constants of the orthorhombic unit cell are $a=5.080$ \AA{} and $b=8.824$ \AA{}. The phonon spectrum calculation confirms its dynamic stability, with no apparent imaginary vibration modes (Figure~S2 in SI). Its space group $C2mm$ preserves the mirror symmetries $m_b$, $m_c$, and a 2-fold rotation symmetry $C_{2a}$, permitting a net polarization along the $a$-axis. As indicated in Figure~\ref{F2}(b), the magnetic sublattices with opposite spins can be connected by the $T\tau$ symmetry operation between the next-nearest-neighbor sites [although the $T\tau$ symmetry is broken between the nearest-neighbor sites as indicated in Figure~\ref{F1}(d)], which enforces the spin degeneracy across the entire Brillouin zone, as shown in Figure~\ref{F3}(a) and Figure~S3 in SI. 

\begin{table}[b]
	\caption{DFT energies, space groups, and band gaps of seven possible magnetic orders in Fe$_3$O$_5$ monolayer. The default $U_{\rm eff}$ is set as $4$ eV for Fe's $3d$ orbitals and more results of other $U_{\rm eff}$'s can be found in Figure~S1 in SI. The SAFM order is taken as the reference, which owns the lowest energy.}
	\begin{tabular*}{0.7\textwidth}{@{\extracolsep{\fill}}cccccc}
		\hline
		&Energy (meV/f.u.) & Space group & Gap (eV)\\
		\hline
		FM & 658.75 & $P\bar{6}2m$ & 0 \\
		SAFM & 0.00 & $C2mm$ & 0.6\\
		ZAFM & 114.01 & $P2_1am$ & 0\\
		YAFM &  141.37 & $P\bar{6}2m$ & 0\\
		UUD-FiM & 58.82 & $Pm$ & 0.2\\
		UUD-AFM & 6.66 & $Pm$ & 0.4\\
		UUD-AFM2 & 174.69 & $Pm$ & 0.7\\
		\hline
	\end{tabular*}
	\label{Tab1}
\end{table}

However, this global $T\tau$ symmetry can be broken in the UUD-AFM state [Figure~\ref{F2}(f)], which is the second-lowest energy one (only $6.66$ meV/f.u. higher). The phonon spectrum of UUD-AFM structure is also dynamically stable (Figure~S4 in SI). Via the spin-lattice coupling, this magnetic order further distorts the crystalline structure to a monoclinic one (space group $Pm$) with $a=5.109$ \AA{}, $b=8.812$ \AA{}, and the interaxial angle $\gamma=90.5^\circ$. Then an additional polarization along the $b$-axis is allowed in the UUD-AFM state. Due to the breaking of $T\tau$ symmetry, the spin-up and spin-down channels are split, as shown in Figure~\ref{F3}(b). 

According to the density of states [Figures~\ref{F3}(a-b)], the occupied electronic states near the Fermi level are predominantly contributed by Fe-$3d$ and O-$2p$ orbitals. Here the nominal average valence of Fe ions is $10/3$. Usually, a nonintegral valence implies partially occupied bands, i.e., metallic behaviors, if the electron distribution is uniform. However, here both the SAFM and UUD-AFM phases exhibit insulating properties according to its electronic structures. This anomalous property lies in the spontaneous charge ordering, i.e., the formation of Fe$^{3+}$-Fe$^{4+}$ order with a ratio of $2:1$. As visualized in Figures~\ref{F3}(c-d), such a charge ordering leads to inhomogeneous distribution of electron density at Fe sites, as well as local magnetic moments ($\sim4.1$ $\mu_{\rm B}$ and $\sim3.6$ $\mu_{\rm B}$ in the ratio of $2:1$).
	
It is noteworthy that the optimal lattice constant $a$ in UUD-AFM is $0.03$ \AA{} larger than that of SAFM, implying that a magnetic phase transition may be possible by applying a uniaxial tensile strain. As shown in Figure~\ref{F3}(e), the phase transition from SAFM to UUD-AFM indeed occurs at $a=5.156$ \AA{} (corresponding to $1.5\%$ tensile strain of the original ground state). In the UUD-AFM state, the integrated local magnetic moment contributed by the Fe sites amounts to $\sim-0.04$ $\mu_{\rm B}$, and the oxygen sites contribute $\sim0.025$ $\mu_{\rm B}$, which partially cancels the Fe moment. The net magnetization, including contributions from Fe sites, O sites, and the interstitial region, is calculated to be zero, as required for a strictly compensated magnetic insulator with a band gap of $\sim0.4$ eV. Such strict compensation is enforced by the full filling of spin-up and spin-down bands, i.e., $N_{\rm \uparrow}$=$N_{\rm \downarrow}$., where $N$ is the integer number of occupied bands. Then, the total net magnetization $M$=$\mu_{\rm B}$($N_{\rm \uparrow}$-$N_{\rm \downarrow}$)=0.
Even considering the spin-orbit coupling (SOC), the net magnetic moment remains negligible, i.e. $\sim0.002$ $\mu_{\rm B}$/Fe in our DFT calculation. Thus, the UUD-AFM phase is a compensated ferrimagnet.

Considering its polar point group $m$ (space group $Pm$), the ferroelectric polarization is allowed along both the $a$- and $b$-axes. According to the Berry phase value difference between the $+P$ and $-P$ states (the $-P$ state is constructed via the spatial inversion operation of $+P$ state), the calculated polarization reaches $12.54$ $\mu$C/cm$^2$ along the $a$-axis and $34.87$ $\mu$C/cm$^2$ along the $b$-axis. These values are exotically large for the type-II multiferroicity. These abnormal results come from the so-called fractional quantum ferroelectricity (FQFE), in which a polarization with fractional multiples of a quantum ($Q$) arises from the fractional lattice period displacements of ions (more precisely the Wannier centers) between symmetry-equivalent sites \cite{Ji2024NC,Xiang2025PRL,He2025PRL,Wu2025PRL}. In our Fe$_3$O$_5$ case, the movements of oxygen ions during the hypothetical ferroelectric switching from $+P$: ($+P_a$, $+P_b$) to $-P$: ($-P_a$, $-P_b$) are almost across the fractional lattice period (Figure~S5 in SI). By deducting $Q/3$ along the $a$-axis and $Q/2$ along the $b$-axis, the residual polarizations are $-0.14$ $\mu$C/cm$^2$ along the $a$-axis and $2.09$ $\mu$C/cm$^2$ along the $b$-axis, which are purely from spin-lattice coupling and within the expected range of magnetostriction \cite{Dong2019NSR}. 

\begin{table}
	\caption{Calculated magnetic moments. $P_a$ and $P_b$ denote the ferroelectric polarization components along the $a$- and $b$-axis, respectively. $m_{\rm Fe}$ represents the Fe-site-projected integrated spin moment summed over all six Fe ions within the default Wigner-Seitz spheres. $m_{\rm \uparrow}$ and $m_{\rm \downarrow}$ denote the corresponding integrated values for the three spin-up and three spin-down Fe ions, respectively, and satisfy $m_{\rm Fe}=m_{\rm \uparrow}+m_{\rm \downarrow}$. $M$ is the net magnetic moments per u.c. with contributions from all ions.}
	\begin{tabular*}{0.7\textwidth}{@{\extracolsep{\fill}}ccccccc}
		\hline 
		&&$m_{\rm{\uparrow}}$ ($\mu_{\rm B}$)&$m_{\rm{\downarrow}}$ ($\mu_{\rm B}$)& $m_{\rm Fe}$ ($\mu_{\rm B}$)&$M$ ($\mu_{\rm B}$)\\
		\hline
		&$+P_a,+P_b$&11.77&-11.81&-0.04&0\\
		&$-P_a,-P_b$&11.81&-11.77&0.04&0\\	
		&$+P_a,-P_b$&11.81&-11.77&0.04&0\\
		&$-P_a,+P_b$&11.77&-11.81&-0.04&0\\
		&$+P_a,0$&11.79&-11.79&0&0\\
		&$-P_a,0$&11.79&-11.79&0&0\\		
		\hline
	\end{tabular*}
	\label{Tab2}
\end{table}

Upon this hypothetical ferroelectric switching, an electron redistribution also occurs, as shown in Figure~S6 in SI. This induces a modulation of the local moments of all Fe ions ($m_{\rm Fe}$) from $-0.04$ $\mu_{\rm B}$ to $0.04$ $\mu_{\rm B}$, as summarized in Table~\ref{Tab2}, which is an indication of  hidden magnetoelectricity. However, the total magnetization ($M$) containing contributions from all ions remains zero, as required by the compensated ferrimagnetism. Thus, the macroscopic magnetoelectric response is recessive.

\begin{figure*}
	\includegraphics[width=0.95\textwidth]{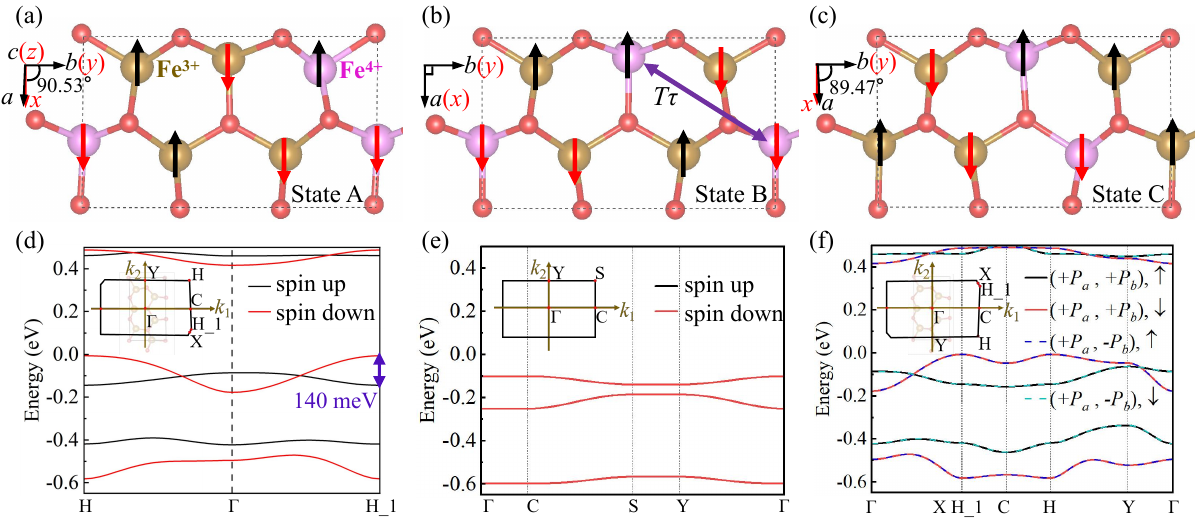}
	\caption{(a-c) Top views of three UUD-AFM states. Fe$^{3+}$ and Fe$^{4+}$ sites are distinguished by colors. (a) State A with ($+P_a$, $+P_b$). (b) State B with ($+P_a$, $0$) (space group $P2_1am$). (c) State C with ($+P_a$, $-P_b$). (d) Spin-dependent electronic band splitting (without SOC) for the state A. (e) Band structure for state B. (f) Reversal of spin-dependent band splitting upon the ferroelectric switching of $\pm P_b$. Insets of (d-f): the corresponding Brillouin zones of (a-c). Note that the high symmetry points of states A and C are not identical, but can coincide after a mirror symmetry operation.}
	\label{F4}
\end{figure*}

Then it is necessary to clarify which component of polarization is primarily responsible for the hidden magnetoelectricity. Instead of directly applying the spatial inversion operation, the magnetic order is tuned by shifting the UUD sequence, as shown in Figures~\ref{F4}(a-c). For such a Fe-trimer unit along the $b$-axis, there are three states: the states A and C have identical energy, which is lower than that of state B by $17.3$ meV/f.u.. The charge ordering pattern is also changed correspondingly. The polarization component along the $b$-axis is $+P_b$ for state A, $0$ for the state B (space group orthorhombic $P2_1am$), and $-P_b$ for state C, while their polarizations along the $a$-component are almost identical. As summarized in Table~\ref{Tab2}, the switching of only the $P_b$ component can switch $m_{\rm Fe}$, while $P_a$ is irrelevant to the magnetoelectric switching. For completeness, $m_{\rm Fe}$ is always zero in the state B with zero $P_b$, regradless of the sign of $P_a$.

Figure~\ref{F4}(d) shows the spin-resolved electronic band structure of state A of UUD-AFM. As expected, the Zeeman-type splitting generally occurs for all bands \cite{yang2023apl162403}. More interestingly, a prominent alternating characteristic is also evidenced along the H-$\Gamma$-H$\_1$ path around the Fermi level, mimicking the behavior of altermagnets. Due to the global Zeeman-type splitting, the spin-up and spin-down bands at the $\Gamma$ point are not degenerated anymore. The maximum value of splitting reaches $140$ meV for the top valence bands. In addition, the 2D color maps of the spin splitting energy $\Delta E(k)$ for three band pairs around the Fermi level have also been provided over the entire 2D Brillouin Zone, as shown in Figure~S7 in SI. Both band pairs I and IV exhibit pronounced alternating behavior, and the spin polarization direction has even parity under the transformation $k\rightarrow-k$. In contrast, band pair II exhibits predominantly Zeeman-type splitting. The splitting of band pair III is weak, with characteristics of both types. All these observations indicate that the system possesses a hybrid spin-splitting mechanism. 

Then it is interesting to ask why the bands exhibit the altermagnetic-like splitting? In fact, even in the absence of lattice distortion, the UUD-AFM order alone in this pyramid structure inherently breaks the global $T\tau$ and $PT$ symmetries. However, when the weak magnetism-induced lattice distortions are neglected, the local $TR$ and $TM$ symmetries between opposite-spin pyramids remain strictly intact, giving rise to the altermagnetic-like behaviors. Since the FQFE arises from the symmetry-allowed Wannier-center displacements, it will not break the $TR$ and $TM$ symmetry of the crystal structure. Only the portion of magnetism-induced polarization can break the $TR$ and $TM$ symmetry, which is a weak effect. Therefore, including such weak distortions only slightly perturbs these local symmetries, and the altermagnetic-like behavior is largely preserved (see Figure~S7 in SI), which is especially prominent for the wide bands near the Fermi level.

For comparison, the band degeneracy is observed in state B of UUD-AFM state, as shown in Figure~\ref{F4}(e), which is protected by the $T\tau$ symmetry between its antiparallel-spin sites as indicated in Figure~\ref{F4}(b). The difference between states A and B suggests that the emergence of band splitting is associated with $P_b$.
Then is it possible to flip the band splitting by switching $P_b$'s direction? However, due to the weak monoclinic distortion of states A and C, their corresponding Brillouin zones are slightly different [see the insets of Figure~\ref{F4}(d) and \ref{F4}(f) for comparison], which can become coincident via a mirror symmetry operation $m_{k_1}$. By mapping the corresponding $k$-paths, the spin-up bands of state A ($+P_b$) coincide with the spin-down bands of state C ($-P_b$), and vice versa, as shown in Figure~\ref{F4}(f). In addition, we have also provided additional material examples. The Mn$_3$O$_5$ monolayer with the UUD-AFM order exhibits ferroelectric switching properties similar to those of Fe$_3$O$_5$, as illustrated in Figure~S8 in SI.

The spin-dependent conductivity is a direct physical consequence of bands' spin-splitting, which is especially useful for narrow-gap semiconductors. For Fe$_3$O$_5$ monolayer, the magnetic point group is $m'$ for the UUD-AFM $\pm P_b$ states, whose symmetry operation $TM_z$ has no restriction on the in‑plane spin-dependent conductivity tensors. The relationship among the spin-dependent conductivity tensors can be directly described as: $\sigma_{ij}^\uparrow\neq\pm\sigma_{ij}^\downarrow$ \cite{tao2025prb224423,yang2025AFMe17921,Jakub2021PRL127701,Ma2021nc2846}, where $\uparrow$ and $\downarrow$ are spin indices, $i$ and $j$ are indices of Cartesian components ($x$ or $y$), as indicated in Figures~\ref{F4}(a-c). Figure S9(a) shows our calculated spin-resolved conductivity as a function of energy with respect to the Fermi level at 5 K. It is observed that $\sigma_{xx}^\uparrow\neq\sigma_{xx}^\downarrow$, $\sigma_{yy}^\uparrow\neq\sigma_{yy}^\downarrow$, and $\sigma_{xy}^\uparrow\neq\pm\sigma_{xy}^\downarrow$, as expected from symmetry analysis.

\begin{figure}
	\includegraphics[width=0.5\textwidth]{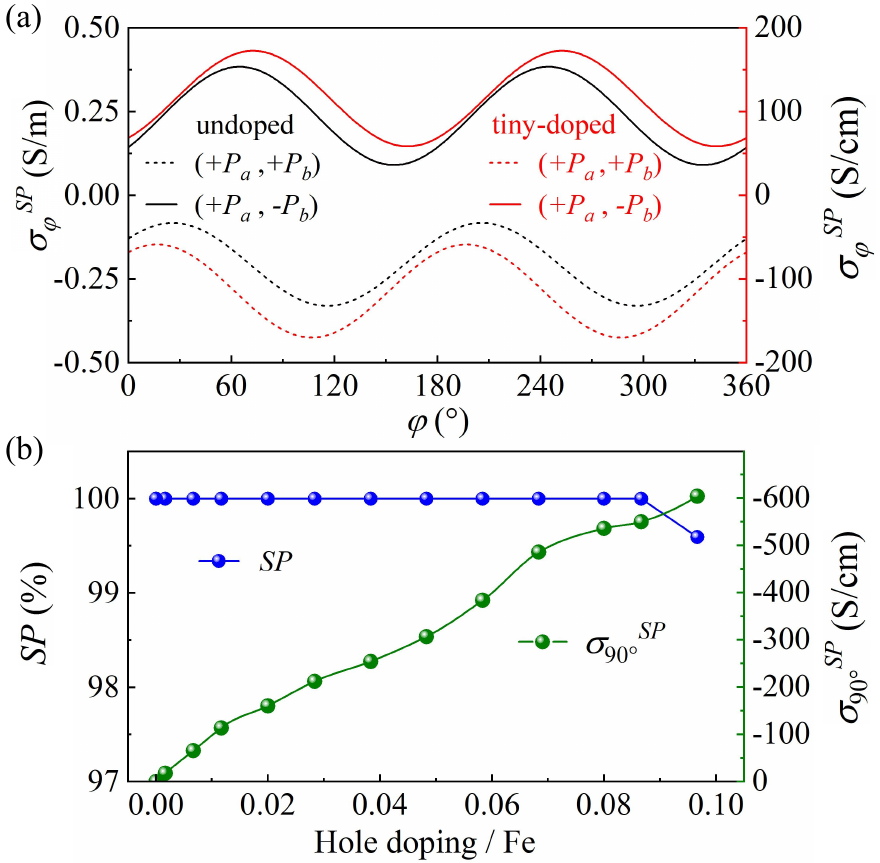}
	\caption{Spin-dependent conductivity of UUD-AFM. (a) Anisotropic $\sigma_{\varphi}^{SP}$ for both undoped and tiny-doped cases at $5$ K. States A and C exhibit a supplementary angle relationship. The hole doping is achieved by setting the Fermi level at $-0.023$ eV, corresponding to $0.02$ holes per Fe. (b) The $\sigma_{90^\circ}^{SP}$ and the corresponding ratio of spin-polarization as a function of hole doping concentration for state A.} 
	\label{F5}
\end{figure}

Then, along arbitrary in-plane direction, the spin-resolved diagonal conductivity can be obtained as: $\sigma_{\varphi}^s=\sigma_{xx}^s\cos^2\varphi+\sigma_{yy}^s\sin^2\varphi+\sigma_{xy}^s\sin(2\varphi)$ \cite{tao2025prb224423,yang2025AFMe17921,Jakub2021PRL127701,Ma2021nc2846}, where $s$ is the spin index and $\varphi$ is the azimuthal angle. The spin-polarized conductivity, defined as $\sigma_{\varphi}^{SP}=\sigma_{\varphi}^\uparrow-\sigma_{\varphi}^\downarrow$, is shown in Figure~\ref{F5}(a) for both undoped and slightly doped cases. It is obvious that $\sigma_{\varphi}^{SP}$ is strongly anisotropic, as a consequence of altermagnetic-like band splitting near the Fermi level. However, in altermagnets, there are always some special directions along which $\sigma_{\varphi}^{SP}$ becomes zero due to the symmetry protection \cite{tao2025prb224423,yang2025AFMe17921}. In contrast, here $\sigma_{\varphi}^{SP}$ is always nonzero, a feature of  compensated ferrimagnets. In addition, $\sigma_{\varphi}^{SP}$ is ferroelectrically switchable, namely $\sigma_{\varphi}^{SP}$ of $+P_b$ is equal to $-\sigma_{180^\circ-\varphi}^{SP}$ of $-P_b$. Such a relationship of supplementary angle originates from $P_b$-switching-induced scalar modulation of spin-dependent conductivity tensors, as shown in Figure~S9(b) in SI.
Such anisotropy and ferroelectric switchability do not qualitatively change upon slight doping, although the value of spin-polarized conductivity increases by several orders of magnitude.

Taking $\varphi=90^\circ$ for example, the spin-polarized conductivity ($\sigma_{90^\circ}^{SP}$) and the ratio of spin-polarization, defined as $SP=\lvert\sigma_{90^\circ}^\uparrow-\sigma_{90^\circ}^\downarrow\lvert/\lvert\sigma_{90^\circ}^\uparrow+\sigma_{90^\circ}^\downarrow\lvert$, have been calculated as a function of hole doping concentration, as shown in Figure~\ref{F5}(b). It is found that the system exhibits spin polarization ratios close to $100\%$ at low temperatures, as well as for other $\varphi$'s. With increasing the hole doping concentration, the value of $SP$ remains considerably large, e.g. above $99\%$ in our calculated region, surpassing the reported values for various altermagnets: RuO$_2$ ($59\%$), FeSb ($50\%$), CuF$_2$ ($78\%$) and K$V_2$Se$_2$O ($60\%$) \cite{Ma2021nc2846,tao2025prb224423,zhangAPL152401}. Such prominent spin polarization ratios of the UUD-AFM Fe$_3$O$_5$ monolayer originate from the Zeeman-type band splitting.

\section{Conclusions}
In summary, based on a model system, Fe$_3$O$_5$ monolayer, we have proposed a mechanism of compensated ferrimagnets with both altermagnetic-type and Zeeman-type splittings. Such a hybrid band splitting is driven by its special UUD magnetic order and charge ordering, which is ferroelectrically switchable. As a narrow-gap semiconductor, Fe$_3$O$_5$ exhibits highly anisotropic spin-polarized conductivity, with the polarization ratio above 99$\%$ under zero net magnetization, which enables high-performance operation as a spin injector and filter. Mediated by magnetoelectric spin-charge coupling, this spin-polarized conductivity is electrically switchable. These findings indicate that our work opens a promising avenue for future studies of spintronic devices and magnetoelectric devices.	

\section{Computational Methods}
First-principles calculations based on density functional theory (DFT) were performed with the projector augmented-wave (PAW) pseudopotentials as implemented in the Vienna {\it Ab initio} Simulation Package (VASP) ~\cite{kresse1996Prb}. The exchange-correlation functional was treated using Perdew-Burke-Ernzerhof (PBE) parametrization of the generalized gradient approximation (GGA) ~\cite{perdew1996Prl}. To avoid layer interactions, a vacuum space of $20$ \AA{} thickness was added along the $c$-axis. The energy cutoff was fixed at $500$ eV. The $\Gamma$-centered $6\times4\times1$ Monkhorst-Pack \textit{k}-mesh was adopted for the unit cell. The convergence criterion for the energy was $10^{-6}$ eV for self-consistent iteration, and the Hellmann-Feynman force was set to $0.01$ eV/\AA{} during the structural optimization. The Hubbard $U$ was applied using the Dudarev parametrization~\cite{dudarev1998Prb}. As reported previously, the correction of $U_{\rm eff}=4$ eV was imposed on Fe's $3d$ orbitals~\cite{nc4702}. Phonopy was adopted to calculate the phonon band structures~\cite{TOGO20151SCR}. Ferroelectric polarization was calculated using the standard Berry phase method~\cite{King1993prb}.The spin-resolved conductivity was calculated within the framework of Boltzmann transport theory using the relaxation time approximation of 10 fs, as implemented in the BoltzWann module~\cite{PIZZI2014CPC} of the WANNIER90 package ~\cite{Pizzi_2020JPCM,MOSTOFI2008Arash}. For the conductivity calculations, the $k$-point mesh of $150\times100\times1$ was used.
	
\begin{acknowledgement}
We thank Jun Chen and Ziwen Wang for useful discussions on the classical Heisenberg spin model. This work was supported by the National Key Research and Development Program of China (No. 2025YFA1411100) and National Natural Science Foundation of China (Grant No. 12325401 and 12274069). This research work was supported by the Big Data Computing Center of Southeast University and the Center for Fundamental and Interdisciplinary Sciences of Southeast University.		
\end{acknowledgement}

\begin{suppinfo}	
	The following files are available free of charge.
	\begin{itemize}
		\item More test results of $U_{\rm eff}$ parameters, phonon spectrum, band structure of SAFM state, 2D color maps of spin-splitting energy projections and spin-resolved conductivity in the UUD-AFM state, details of ferroelectric switching from $+P$ to $-P$ states, the properties of Mn$_3$O$_5$ monolayer.
	\end{itemize}
	
\end{suppinfo}
	
\bibliography{reference}
\end{document}